\documentclass[9pt,twocolumn,twoside]{article}
\usepackage[outdir=./]{epstopdf}
\usepackage{subcaption}
\usepackage{cleveref}
\usepackage{graphicx}

\usepackage{bm}

\usepackage{authblk}

\title{Machine learning based compact photonic structure design for strong light confinement}

\author[1,5,*]{MIRBEK TURDUEV}
\author[2,5]{CAGRI LATIFOGLU}
\author[3,6]{IBRAHIM HALIL GIDEN}
\author[4,5]{Y. SINAN HANAY}

\affil[1]{Department of Electrical and Electronics Engineering, TED University, Ankara 06420, Turkey}
\affil[2]{Department of Industrial Engineering, TED University, Ankara 06420, Turkey}
\affil[3]{Department of Electrical and Electronics Engineering, TOBB University of Economics and Technology, Ankara 06560, Turkey}
\affil[4]{Department of Computer Engineering, TED University, Ankara 06420, Turkey}
\affil[5]{Photonic Networks Research Group, TED University, Ankara 06420, Turkey}
\affil[6]{Nanophotonics Research Group, TOBB University of Economics and Technology, Ankara 06560, Turkey}
\affil[*]{Corresponding author: mirbek.turduev@tedu.edu.tr}

\date{Compiled \today}

\begin{document}
\maketitle
\begin{abstract}
  We present a novel approach based on machine learning for designing photonic structures. In particular, we focus on strong light confinement that allows the design of an efficient free-space-to-waveguide coupler which is made of Si- slab overlying on the top of silica substrate. The learning algorithm is implemented using bitwise square Si- cells and the whole optimized device has a footprint of $\boldsymbol{2 \, \mu m \times 1\, \mu m}$, which is the smallest size ever achieved numerically.  To find the effect of Si- slab thickness on the sub-wavelength focusing and strong coupling characteristics of optimized photonic structure, we carried out three-dimensional  time-domain numerical calculations. Corresponding optimum values of full width at half maximum and coupling efficiency were calculated as  $\boldsymbol{0.158 \lambda}$ and $\boldsymbol{-1.87\,dB}$ with slab thickness of $\boldsymbol{280nm}$. Compared to the conventional counterparts, the optimized lens and coupler designs are easy-to-fabricate via optical lithography techniques, quite compact, and can operate at telecommunication wavelengths. The outcomes of the presented study show that machine learning can be beneficial for efficient photonic designs in various potential applications such as polarization-division, beam manipulation and optical interconnects. 
\end{abstract}

  Studies on subwavelength light focusing phenomenon are increasing over time due to immense needs of light energy enhancement via confining it to a narrow region \cite{Betzig1991, Pendry2000, NicholasFangHyesogLeeChengSun2005}. Light focusing into a tight spot smaller than the wavelength of light may find place in various optical applications including nanolithography, optical microscopy, optical measurements and optical data storage. Especially, high-resolution imaging and strong beam coupling can be achieved by using the light focusing effect beyond the diffraction limit \cite{Jacob2006,Poddubny2013}. Various subwavelength focusing lenses have already been developed using metasurfaces \cite{Yu2014}, plasmonics \cite{Ye2011}, metamaterials \cite{Kannegulla2016} and all-dielectric materials \cite{bor2016differential} as well as exploiting super-oscillation \cite{Huang2009} and hypergrating techniques \cite{Thongrattanasiri2009}. 
  
  Conventional theoretical approaches may not fully provide analytical solutions for complex photonic structures. Furthermore, theoretically designed systems may not satisfy the feasibility and performance requirements such as compactness, efficiency, bandwidth and energy loss drawbacks.  For this reason, different types of optimization algorithms such as evolutionary algorithm \cite{bor2016differential}, inverse optimization \cite{Lu2013}, topology optimization \cite{Borel2004}, nonlinear search algorithm \cite{Shen2014}, and the direct binary search algorithm \cite{Shen2015} have been implemented to design efficient photonic integrated structures. However, to the best of authors’ knowledge, this is the first time that machine learning algorithm is utilized to design photonic devices with desired optical properties. 
  
  Machine learning has attracted much attention from both academia and industry recently. Researchers leveraged machine learning successfully in developing a self-learning robot which can withstand failures \cite{Cully2015a}, predicting hepatitis B positive patients \cite{Ye2003a}, improving components' performances in optical communication systems \cite{Zibar2016}, detecting malicious software automatically \cite{Rieck2011a} and understanding ecological phenomena \cite{Olden2008a}. Well-known applications in industry include driverless cars, language translation services (e.g. Google translate), credit decision systems and anti-virus software. In all those diverse examples, the power of machine learning lies in accurate modeling and characterization of complex relationships within the systems.
  
  \begin{figure*}[!t]
    \centering
    \includegraphics[width=\linewidth]{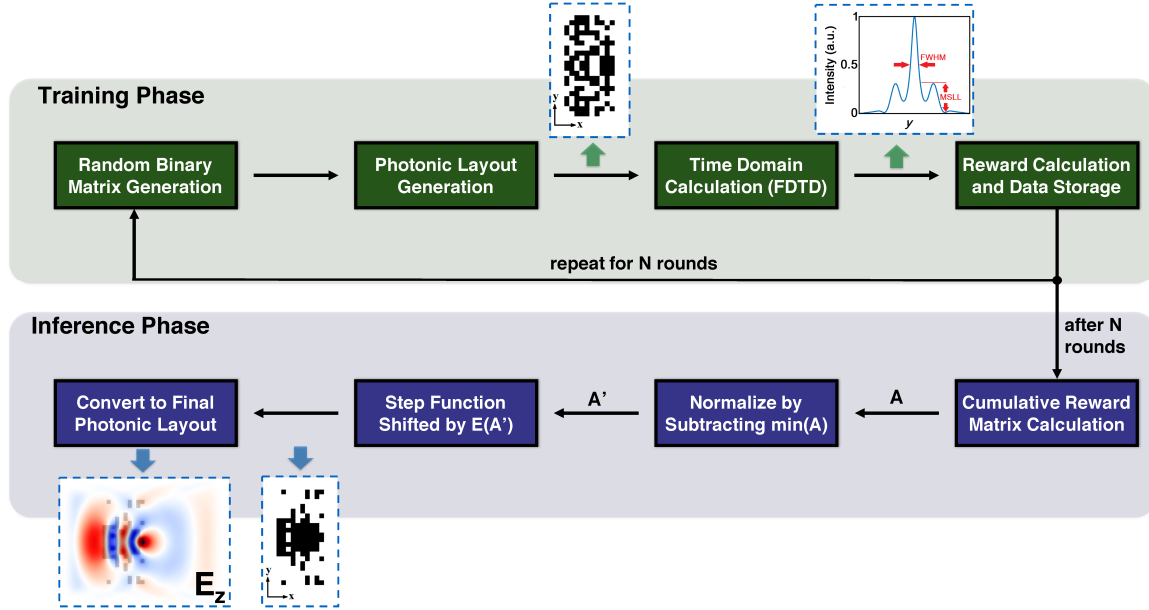}
    \caption{Flowchart of ARLA optimization algorithm for subwavelength light focusing purpose.}
    \label{flowchart}
  \end{figure*}
  
  In this study, we propose using a machine learning algorithm to construct highly efficient and compact focusing lenses that are able to surpass the light diffraction limit. The algorithm we use is called Additive Reinforcement Learning Algorithm (ARLA) \cite{Cagri2017} and like all the machine learning algorithms, ARLA consists of two phases: training and inference. In the training phase, which consists of a number of rounds, ARLA generates random photonic structures, and records the optical performance of each random structure. Later in the inference phase ARLA uses the information collected in the training phase to come up with a final layout. 
  
  The details of ARLA can be found in \cite{Cagri2017}, which combines the additive updates feature of the perceptron algorithm \cite{Rosenblatt1958} and the reward for state idea of reinforcement learning \cite{Sutton1998}. However, we briefly explain the steps in ARLA as shown in Figure \ref{flowchart}. In the upper diagram of the figure, operations in a single round are depicted. The training phase starts with creating a random photonic structure that needs to be mirror-symmetric along x-axis to obtain the focal point at the optical axis. That half of the structure is represented with a 10x10 binary matrix, where a “1” corresponds to Si- and “0” corresponds to air cells of 100nm x 100nm. Thus, the size of complete photonic structure equals $2 \, \mu m \times 1\, \mu m$. Next, finite-different time-domain (FDTD) method is used to calculate full-width at half maximums (FWHM) and maximum side lobe levels (MSLL) of the focused light beam using MEEP \cite{Oskooi2010}. Here, FWHM and MSLL values are arranged as inputs for the reward function, which is defined as: 
  \begin{equation}
    R=R_{max}-(w_1\times FWHM+w_2 \times MSLL).
  \end{equation}
  
  In this expression, $w_1$ and $w_2$ represent weights that control the trade-off between FWHM and MSLL values. The objective of the design problem is to find a photonic layout that minimizes both FWHM and MSLL; when both FWHM and MSLL values approaches to zero, reward approaches to $R_{max}$, which is an arbitrarily chosen constant that satisfies the following condition: 
  
  \begin{equation}
    R_{max} > (w_1\times FWHM+w_2 \times MSLL).
  \end{equation}

  In the second phase, the inference phase, ARLA calculates a cumulative reward matrix from the results obtained in the training phase. A round matrix is multiplication of the binary matrix evaluated and its reward. Then ARLA normalizes this matrix $A$ by subtracting the minimum value from all the elements. Next, ARLA applies a step function, which is shifted by the mean of $A^\prime$, to convert the normalized  matrix $A^\prime$ to a binary matrix. Finally, the binary matrix is converted back to a photonic structure, and this final structure is the final output of the algorithm, as can be seen from the inset in the figure.   
  
  The number of training rounds N depends on the desired focusing performance and the size of the photonic structure. In our study that number was chosen empirically. Due to independent nature of the training rounds, ARLA can be fully parallelizable. We distributed the training rounds to 12 different virtual machines each having two virtual CPUs and 3.75 GB RAM. ARLA can work with larger matrices as well, at the cost of increased training times. 
  
  \begin{figure*}[!th]
    \centering
    \includegraphics[width=\linewidth]{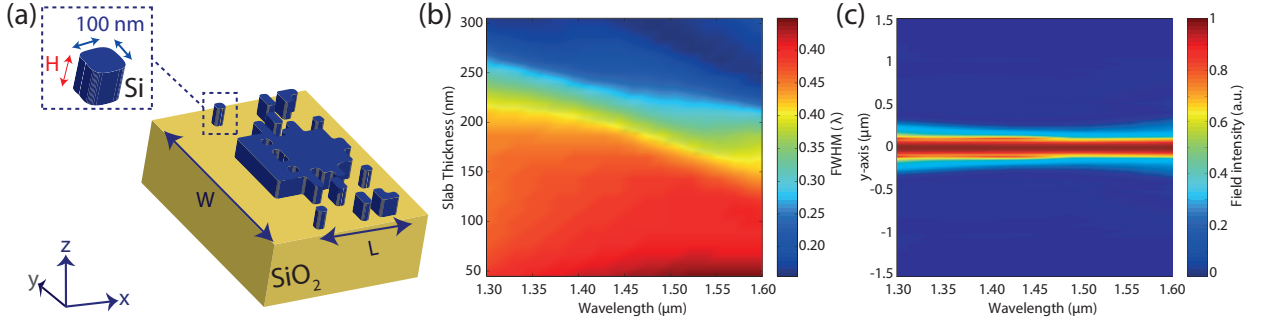}
    \caption{(a) Optimized bitwise photonic structure design as a near field sub-wavelength lens. (b) FWHM map depending on the slab thickness and incident wavelength variations. (c) The cross-sections of field intensities are taken at the output and plotted as a map depending on the incident wavelengths.}
    \label{fig2}
  \end{figure*}
  
  The output of ARLA gives an optimized structure with the highest reward providing the subwavelength focusing effect, which is schematically drawn as inset in Fig. \ref{flowchart}. The dielectric medium is optimized in bitwise scale via the machine learning ARLA method in order to control the near-field strong focusing effect, see the corresponding steady-state electric field $E_z$ distribution in the same figure. Obtained subwavelength focusing effect through the optimized scattering structure can be related to constructive interference of propagating waves \cite{Wiersma2013}: The optimized “turbid” structure is perturbed in bitwise-scale so that nanoscale local modifications exist in the corresponding refractive index distribution. The incident wave, then, encounters multiple light scattering while propagating through the structure. Such scattering mechanism enhances the constructive interferences at the output so that a strong light focusing may emerge \cite{Mosk2012}.
  
  The optimized lens design is constructed on a silicon-on-insulator (SOI) substrate and composed of Si- slab having the refractive index of $n_{Si}=3.47$, see Fig. \ref{fig2}(a). Corresponding geometrical parameters of the lens structure are represented as an inset in the same figure. The width and height of photonics lens are equal to $\{W,L\}=\{2\,\mu m,1\,\mu m\}$. 3D FDTD calculations are carried out for the optimized structure in order to analyze the effect of Si- slab thickness on the subwavelength focusing performance \cite{lumerica}. FWHM values are calculated at the output face of the optimized design with respect to the variations of slab thicknesses H and the incident wavelengths with transverse magnetic-like (TM-like) polarization. This relationship is plotted as a FWHM map in Fig. \ref{fig2}(b). The calculated map shows that the excited light focuses with FWHM values below $0.20\lambda$ while the slab thickness to be in the range of 250nm-300nm. Furthermore, the minimum FWHM value was calculated as $0.155\lambda$, which implies that the diffraction limit of light could be defeated in the near-field by the help of the optimized design. To be compatible with SOI technology and optical communication systems, the slab thickness of the optimized structure is fixed to H=280 nm. 
  
\begin{figure}[!b]
    \centering
    \includegraphics[width=\linewidth]{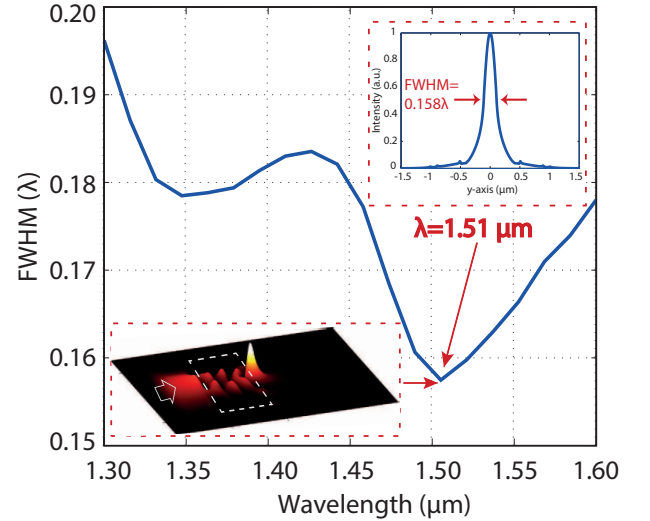}
    \caption{ Calculated FWHM values depending on incident light wavelengths of the photonic slab lens having $H=280nm$ thickness. 
      Steady state electric field profile with its transverse cross-sectional  intensity profile at $\lambda={1.51\mu m}$ are given as insets.}
    \label{FWHM}
  \end{figure}

  In order to show the broadband subwavelength focusing effect of the designed photonic lens, corresponding transverse cross-sections of output electric field intensities were calculated for varying incident wavelengths ranging between $1.3 \mu m-1.6 \mu m$ and superimposed in Fig. \ref{fig2}(c) as a transverse cross-sectional intensity map. It should be noted that, while dealing with the strong light focusing effect, it is crucial to consider the emergence of side lobes. Excessive side lobe radiations cause the reduction of the main lobe, which is undesirable for strong light focusing. Therefore, a main purpose of using machine learning is to suppress the intensity levels of the MSLLs besides of minimizing the FWHM values at the focal point. It can be clearly observed in the Fig. \ref{fig2}(c) and Fig. \ref{FWHM} (see inset plot) that incident beam is concentrated into a narrower spot with almost negligible levels of additional undesired side lobes.
 
    \begin{figure*}[!t]
    \centering
    \includegraphics[width=\linewidth]{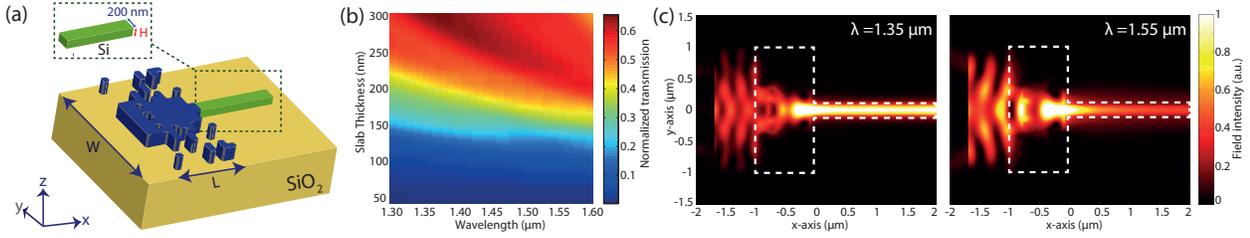}
    \caption{(a) Optimized photonic structure design for air-to-waveguide coupling application. (b) Corresponding transmission efficiency map depending on the slab thickness and incident wavelength variations. (c) Spatial field intensity distributions of optimized structure calculated for the incident wavelengths of $\lambda = 1.35\mu m$ (left) and $\lambda = 1.55 \mu m$ (right). The dashed boxes represent the structure boundaries.}
    \label{optimization}
  \end{figure*}
  
  To quantify the broadband subwavelength focusing effect of the optimized lens structure, the slab thickness is fixed to H=280 nm and the FWHM values of the focused beam are calculated within telecom wavelengths. Corresponding FWHM plot is represented in Fig. \ref{FWHM}. As can be seen from the figure, calculated FWHM values were less than $0.197\lambda$, where the minimum value of FWHM was calculated as $0.158\lambda$ at operating wavelength $\lambda={1.51\mu m}$. The spatial electric field intensity profile with its transverse cross-sectional plot is calculated at $\lambda={1.51\mu m}$ and given as insets in the same figure. The FWHM plot shows that the optimized structure provides a broadband subwavelength focusing effect under $\lambda /6$ within the telecom wavelengths. Such strong subwavelength focusing property as well as small footprint size imply that the optimized photonic lens can also be considered as an alternative and promising solution for free-space-to-waveguide optical coupling applications. In addition, the optimized structure is all-dielectric so that such a photonic device is free of absorption losses and it may operate in a broadband regime comparing to its metamaterial and plasmonic counterparts \cite{turduev2016focusing, Nalimov2013}. The coupling effect is also investigated in detail in the study using proposed machine learning based photonic lens.

  The optimized photonic lens device can also be applied for free-space-to-waveguide coupling problems. For this reason, Si- bulk waveguide with the width of 200nm is butt-coupled to the back face/output of the structure.  3D representation of the optimized free-space-to-waveguide coupler is represented in  Fig. \ref{optimization}(a). The thicknesses of Si- slab and coupled waveguide are scanned in the range of 50nm-300nm to analyze the coupling efficiency depending on the slab thickness. Free-space-to-waveguide coupling efficiency of the optimized structure is plotted according to the variation of incident wavelengths' and given as a transmission map in Fig. \ref{optimization}(b). As can be seen from the map, the coupling efficiency of above $50\% (-3dB)$ is obtained in the case of the slab thickness greater than 250nm. 
  
  In the case of setting the coupler thickness to 280nm, the corresponding coupling efficiency was calculated to be around $65\% (-1.87dB)$ within S (short wavelengths) and C (erbium window) wavelengths. Fig. \ref{optimization}(c) demonstrates steady-state spatial intensity profiles of the optimized coupler having 280 nm thickness for the incident wavelengths of $\lambda=\{1.35\mu m,1.55\mu m\}$. The optimized coupler and the output waveguide are shown as dashed boxes in the figure. It can be inferred from the field intensity calculations in Fig. \ref{optimization}(c) that the proposed coupler design strongly confines the light to the output waveguide with ten times smaller width. In other words, the designed coupler transfers light energy from $2\mu m$ width waveguide to 200nm waveguide (with beam compressing ratio of 10:1) with $65\%$ coupling efficiency.
  
  In conclusion, all-dielectric subwavelength focusing lens and air-to-waveguide coupler were designed utilizing machine learning for the first time. The performance of optimized photonic structure was analyzed employing 3D FDTD numerical calculations. The footprint of whole optimized system is $2\,\mu m\times 1\,\mu m$, which is thought to be very compact compared to the other systems in the literature. A minimum FWHM value of $0.158 \lambda$ and a maximum coupling efficiency of $-1.87dB$ were calculated for the optimized photonic structure with the slab thickness of 280nm. The obtained sub-wavelength light focusing and strong coupling effects can be can be useful in various application domains such as medicine, optical communication, sensing structures and high precision optical device designs.
  
  {\bf Funding.} The authors acknowledge the partial financial support of NATO SPS research grant No: 985048.
  
  {\bf Acknowledgement.} The authors would like to thank Emre Bor for his helpful comments during the optimization process. 
  \bibliographystyle{unsrt}
  \bibliography{library}

\end{document}